\documentclass[runningheads]{llncs}
\usepackage{graphicx}
\usepackage{listings}
\usepackage{array} 
\usepackage{ragged2e}
\usepackage{enumitem}
\usepackage{todonotes}
\usepackage{url}
\usepackage{hyperref}
\usepackage{wrapfig}
\usepackage{subcaption}
\usepackage{caption}
\usepackage{multirow}
\usepackage{flushend}
\usepackage{tikz}
\usepackage{censor}
\usepackage[most]{tcolorbox}
\usepackage{soul}

\usepackage{textcomp}
\usepackage{amsmath}
\usepackage{array, multirow, bigdelim, makecell, booktabs} 

\tcbuselibrary{theorems}
\newtcolorbox{mybox}[1]{colback=white!5!white,colframe=gray!45!white, title =#1,coltitle=black!20!black}

\newcommand{\fig}[1]{Fig.~\ref{fig:#1}}   
  
\newcommand{\tab}[1]{Table~\ref{tab:#1}}

\colorlet{punct}{red!60!black}
\definecolor{background}{HTML}{EEEEEE}
\definecolor{delim}{RGB}{20,105,176}
\colorlet{numb}{magenta!60!black}

\lstdefinelanguage{json}{
    basicstyle=\normalfont\ttfamily,
    numbers=left,
    numberstyle=\scriptsize,
    stepnumber=1,
    numbersep=8pt,
    showstringspaces=false,
    breaklines=true,
    frame=lines,
    backgroundcolor=\color{background},
    literate=
     *{0}{{{\color{numb}0}}}{1}
      {1}{{{\color{numb}1}}}{1}
      {2}{{{\color{numb}2}}}{1}
      {3}{{{\color{numb}3}}}{1}
      {4}{{{\color{numb}4}}}{1}
      {5}{{{\color{numb}5}}}{1}
      {6}{{{\color{numb}6}}}{1}
      {7}{{{\color{numb}7}}}{1}
      {8}{{{\color{numb}8}}}{1}
      {9}{{{\color{numb}9}}}{1}
      {:}{{{\color{punct}{:}}}}{1}
      {,}{{{\color{punct}{,}}}}{1}
      {\{}{{{\color{delim}{\{}}}}{1}
      {\}}{{{\color{delim}{\}}}}}{1}
      {[}{{{\color{delim}{[}}}}{1}
      {]}{{{\color{delim}{]}}}}{1},
}

\definecolor{eclipseStrings}{RGB}{42,0.0,255}
\definecolor{eclipseKeywords}{RGB}{127,0,85}
\colorlet{numb}{magenta!60!black}

\newcounter{theo}[section]
\renewcommand{\thetheo}{\arabic{section}.\arabic{theo}}

\definecolor{airforceblue}{rgb}{0.36, 0.54, 0.66}
\definecolor{amber}{rgb}{1.0, 0.75, 0.0}
\definecolor{antiquebrass}{rgb}{0.8, 0.58, 0.46}
\definecolor{bleudefrance}{rgb}{0.19, 0.55, 0.91}
	\definecolor{bittersweet}{rgb}{1.0, 0.44, 0.37}
	\definecolor{bondiblue}{rgb}{0.0, 0.58, 0.71}

\colorlet{mylinkcolor}{bondiblue}
\colorlet{mycitecolor}{bittersweet}
\colorlet{myurlcolor}{bondiblue}

\makeatletter
\newcommand\footnoteref[1]{\protected@xdef\@thefnmark{\ref{#1}}\@footnotemark}
\makeatother

\colorlet{punct}{red!60!black}
\definecolor{background}{HTML}{EEEEEE}
\definecolor{delim}{RGB}{20,105,176}
\colorlet{numb}{magenta!60!black}

\usepackage{hyperref}
\hypersetup{
    colorlinks=true,
    linkcolor=blue,
    filecolor=blue,      
    urlcolor=blue,
    citecolor=blue
    }

\begin{document}
%


\title{The Microservice Dependency Matrix}





%
\titlerunning{The Microservice Dependency Matrix}
%
\author{Amr S. Abdelfattah\inst{1}\orcidID{0000-0001-7702-0059} \and Tomas Cerny\inst{2}\orcidID{0000-0002-5882-5502}}
%
\authorrunning{Abdelfattah and Cerny}
%

\institute{Computer Science, Baylor University, One Bear Place 97141 Waco, TX, USA \\
\email{amr\_elsayed1@baylor.edu} \and
Systems and Industrial Engineering, University of Arizona, Arizona, USA \\
\email{tcerny@arizona.edu}} 


%
\maketitle              
\begin{abstract}
Microservices have been recognized for over a decade. They reshaped system design enabling decentralization and independence of development teams working on particular microservices. While loosely coupled microservices are desired, it is inevitable for dependencies to arise. However, these dependencies often go unnoticed by development teams. As the system evolves, making changes to one microservice may trigger a ripple effect, necessitating adjustments in dependent microservices and increasing maintenance and operational efforts. Tracking different types of dependencies across microservices becomes crucial in anticipating the consequences of development team changes. This paper introduces the Endpoint Dependency Matrix (EDM) and Data Dependency Matrix (DDM) as tools to address this challenge. We present an automated approach for tracking these dependencies and demonstrate their extraction through a case study.


\keywords{Microservice Dependency \and Static Analysis \and Service Dependency \and System Evolution \and Automated Reasoning}
\end{abstract}

\section{Introduction}

Microservice Architecture is widely used for complex systems that require selective scalability or the decomposition of complex organizational structures into smaller, independently managed units handled by separate development teams. As software systems evolve due to market demands, technological shifts, patches, or optimizations, new features are implemented, and bugs are fixed, potentially introducing new services and system dependencies~\cite{cerny2022microservice}.
Isolated modifications of individual services typically do not cause disruptions to others \cite{microservice2014lewis}. Nevertheless, as systems undergo evolution and dependencies naturally emerge within the architecture, posing challenges to the system's consistency and maintainability. Hence, it becomes crucial to proactively monitor and uphold the principles of low coupling and minimize dependencies within the architecture. In fact, consider a scenario where a critical bug is identified in a particular microservice. By accurately tracking the system dependencies, developers can confidently modify and debug the specific microservice without worrying about unintended consequences or unintended disruptions to other interconnected services. This highlights the importance of actively managing and preserving a low-coupling architecture to ensure the long-term stability and scalability of microservice-based systems.

Recent studies highlight the lack of methods to prevent maintainability problems in microservices~\cite{apolinario2021method}. While existing metrics focus on direct dependencies introduced through endpoint calls between microservices, other aspects introduce dependencies too. For example, the presence of a common data model between microservices can lead to inconsistencies and coupling, where changes in one microservice may require modifications in others. This perspective provides another dimension to understanding the interconnectivity between~microservices.

The main objective of this paper is to introduce and identify system dependencies at different perspectives, including direct endpoint calls and data dependencies, by analyzing the source code of microservices-based systems. We aim to offer a comprehensive understanding of service dependencies.

One of the key contributions of this paper is the development of an automated approach that extracts this dependency information directly from the codebase, ensuring that the obtained insights are up-to-date and free from outdated or stale information. The paper's contributions are summarized as follows:
\vspace{-0.7em}
\begin{itemize}
    \item Describing automated approaches for constructing the Endpoint Dependency Matrix (EDM) and Data Dependency Matrix (DDM) of microservice-based systems.
    \item Implementing a prototype that applies the proposed approaches.
    \item Conducting a case study on a real public microservice project to generate the dependency matrices and discuss the results.
\end{itemize}
\vspace{-0.7em}
The paper is organized as follows. Section~\ref{sec:method} presents the proposed method for constructing the dependency matrices. Section~\ref{sec:cs} presents the case study results for validation. Section~\ref{sec:discussion} discusses the approach and potential threats to validity. Section~\ref{sec:rw} introduces related works. Finally, Section~\ref{sec:conclusion} concludes the paper.

\section{The Proposed Dependency Methodology} \label{sec:method}


The proposed method focuses on capturing the dependencies within microservice systems by considering both endpoints and data entities. Microservices systems utilize specialized frameworks to streamline the development of diverse capabilities. These frameworks often leverage object-oriented concepts and offer robust implementations. Through the utilization of static analysis techniques applied to the source code of the microservices, the necessary components are extracted to facilitate a comprehensive understanding of the system's dependencies.

To construct the EDM, the method identifies the direct endpoint calls within the source code, capturing the dependencies between microservices. The DDM is generated to represent the dependencies based on the shared data entities among microservices. By combining the information from EDM and DDM, a holistic depiction of the system's dependencies is achieved, providing insights into the flow of dependencies between both endpoints and data entities.

This approach serves as a valuable tool for practitioners to gain a comprehensive understanding of the intricate dependencies within microservice systems. By examining the system from both the endpoints and data perspectives, potential bottlenecks, inefficiencies, or critical dependencies can be identified, enabling better decision-making for system maintenance and evolution.




\vspace{-2em}
\begin{figure}[h!]
\centering
\includegraphics[width=0.6\textwidth]{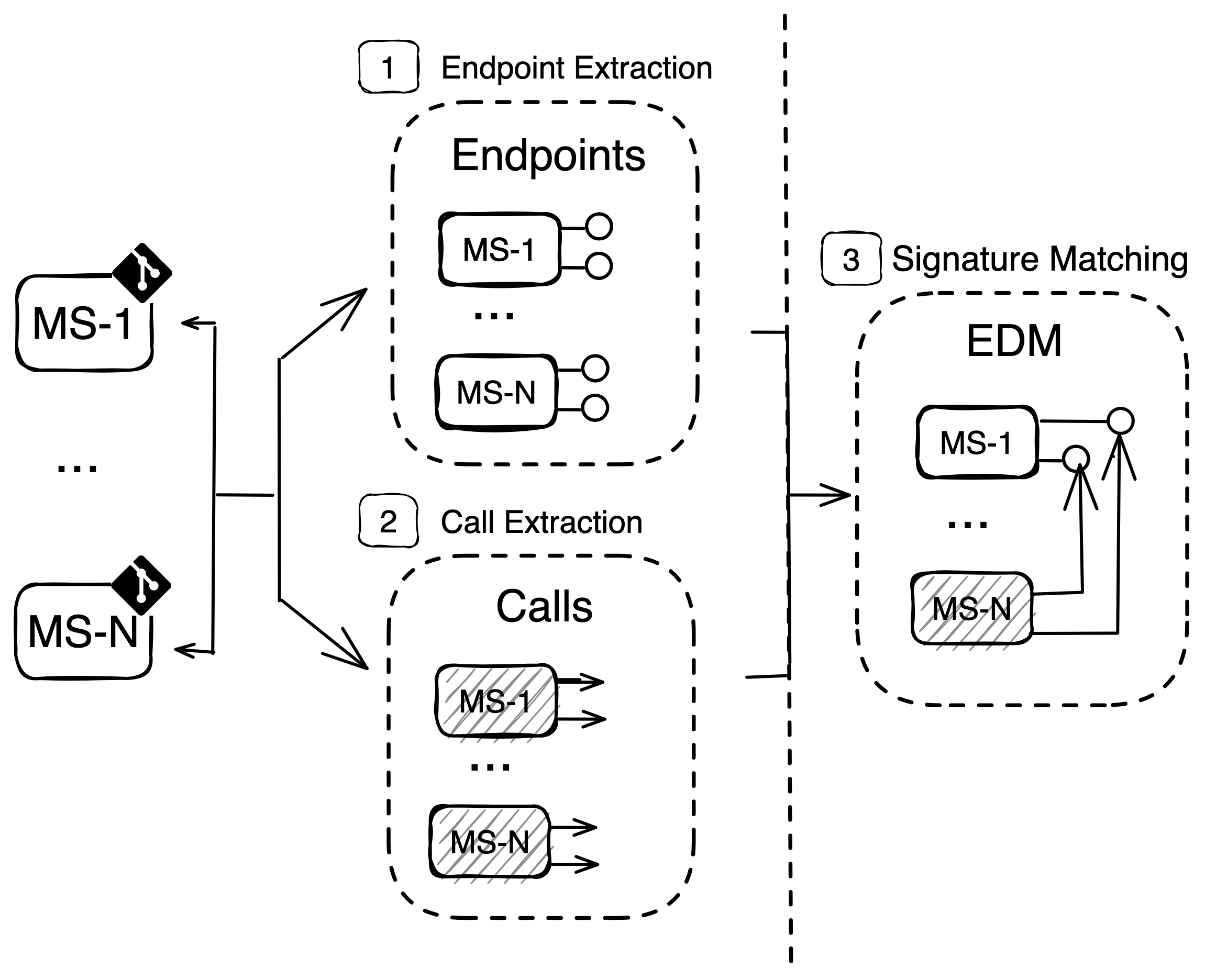}
\vspace{-1em}
\caption{Endpoint Dependency Matrix Generation Process.}
\label{fig:service-view-process}
 \vspace{-3.5em}
\end{figure}

\subsection{Endpoint Dependency Matrix (EDM)}


The dependency between endpoints reveals the interdependencies among different microservices, where one microservice's source code contains a request call to an endpoint of another microservice. Our process examines the distributed source codes of microservices to extract the defined HTTP endpoints and request calls. This process consists of three phases, as depicted in~\fig{service-view-process}: Endpoint Extraction, Call Extraction, and Signature Matching.

In the \textbf{Endpoint Extraction} phase, we identify and extract the HTTP endpoints defined in the source code. Typically, endpoints are specified using framework-specific functions or annotations. This approach ensures consistency in metadata identification. During this phase, we collect various attributes for each endpoint, including the path, HTTP method, parameters, and return type.

The \textbf{Call Extraction} phase focuses on extracting the requests made from the source code. By identifying the corresponding client, we determine where these endpoints are called from other services. Through code analysis, we can gather metadata about every call in the system by identifying the appropriate function call formats specific to the known HTTP library. Therefore, we extract the path, HTTP method, parameters' values, and the expected return type.



The \textbf{Signature Matching} phase involves comparing endpoint method signatures with data and parameters exchanged during REST call interactions. This process finds the matches between endpoint and request calls in the distributed source code. The collected endpoint and call details are merged to establish associations between calls and their corresponding endpoint components. However, direct matching is complex due to the endpoint definition including parameter data types, while request calls involve parameter values or variables in the request's body or path. Our approach initially considers path and parameter count matching. Subsequently, regular expressions are employed to identify the optimal match for parameter types with values in the calls. A successful match signifies a communication path between microservices via the matched endpoint.

Consequently, we can generate an EDM that illustrates the number of request calls between each pair of microservices in the system, thereby displaying the communication dependencies.

\vspace{-1em}
\subsection{Data Dependency Matrix (DDM)}

\begin{figure}[h!]
\centering
\includegraphics[width=0.75\linewidth]{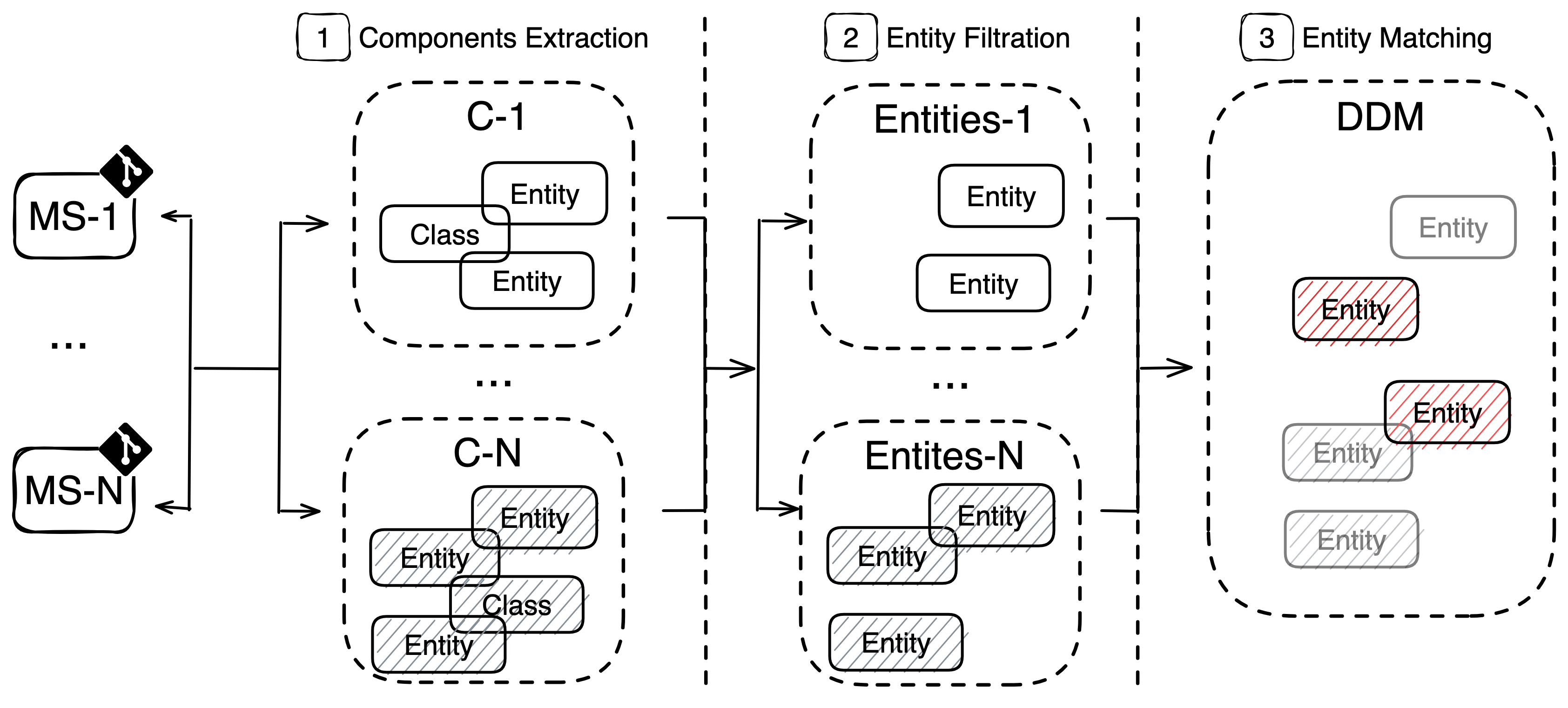}
\vspace{-1em}
\caption{Data Dependency Generation Process.}
\label{fig:context-map-process}
 \vspace{-2em}
\end{figure}

Each microservice establishes a data-bounded context that defines the scope where its specific domain model applies. To identify data dependencies, this method employs static analysis techniques to extract bounded contexts from each microservice's source code. It then proceeds to determine the correspondence between data entities across the individual bounded contexts. The construction process for data dependencies consists of three phases, as illustrated in~\fig{context-map-process}: Components Extraction, Entity Filtration, and Entity Matching.

In the \textbf{Components Extraction} phase, all local classes declared in the project are extracted. Once these classes are identified, the \textbf{Entity Filtration} phase follows, which selects both Data Transfer Objects (DTOs) and classes representing persistent data. It focuses solely on data-related entities, excluding other classes like those serving as REST controllers or internal services. These two phases leverage enterprise standards and frameworks' components, such as annotation descriptors, to differentiate between class types based on their semantic purpose.

\vspace{-2em}

Finally, the \textbf{Entity Matching} phase examines all extracted entities across the microservices to generate a matching list between them. Different bounded contexts may have distinct intentions for the shared entities, resulting in potential variations in the fields they retain. This phase matches entities based on their names, considering if they are the same or similar. Additionally, it examines whether some of their fields share the same data type and possess similar or identical names. This process yields the DDM, which provides insight into the common data entities among microservices.

\section{Case Study} \label{sec:cs}

To demonstrate the effectiveness of our method, we apply it to a real-world scenario. It showcases the capabilities of capturing and understanding the dependencies present in microservice systems. Additionally, we seek to provide valuable insights into the interconnectedness of endpoints and data dependencies, leading to a comprehensive understanding of the system's overall dependency landscape.

Our approach was implemented into a prototype, which we utilized to analyze a publicly available testbench. This allowed us to construct matrices depicting the dependencies of endpoints and data. The comprehension of system dependencies provided by these matrices serves as a valuable tool for facilitating seamless modifications and preserving the maintainability of the system. Moreover, these matrices play a crucial role in monitoring the evolution of dependencies throughout system changes. By generating and analyzing the dependency matrix, developers can track the impact of each commit on the system, observe how it affects system dependencies, and evaluate system coupling and stability. This enables informed decision-making and proactive management of dependencies, ultimately leading to a more robust and adaptable system architecture.






\vspace{-1em}
\subsection{Prototype Implementation and Testbench}

We developed a prototype\footnote{\label{foot-prototype}Prototype: \url{https://github.com/cloudhubs/graal-prophet-utils}} implementation of our proposed approach specifically designed for analyzing Java-based microservices projects utilizing the Spring Boot framework. The prototype utilizes Graal~\cite{duboscq2013graal}, the runtime system developed by Oracle Labs. The prototype takes a GitHub repository containing microservices-based projects as input. It downloads the repository and generates a list of directories for each microservice project.

For the \textbf{Endpoint Dependency Matrix}, the prototype scans the project files for JAX-RS annotations that define endpoints. By combining class-level and method-level annotations, it creates a comprehensive definition for each endpoint, including its path, HTTP method, parameters, and return type. The prototype also scans each microservice to identify the Spring Boot REST client (RestTemplate client) and detects HTTP calls between services. It then applies the signature matching technique to match the detected calls with the corresponding endpoints. The prototype generates a JSON structure that represents the dependencies between microservices and the matched calls. Each microservice name serves as a key in the structure, containing two list values: Dependencies and Dependant services. These lists provide detailed information about the involved endpoints associated with each microservice node.

Regarding the \textbf{Data Dependency Matrix}, the prototype extracts all local classes in the project using a source code analyzer. It filters this list down to classes serving as data entities using persistence annotations (JPA standard entity annotations such as @Entity and @Document). It also considers annotations from Lombok\footnote{\label{foot-lombok}Lombok: \url{https://projectlombok.org}}, a tool for automatically creating data entity objects (e.g., @Data), although these annotations do not explicitly indicate persistence. The prototype then examines the entities of different bounded contexts and their fields, applying the matching rules described above. To detect name similarity, the prototype employs the WS4J\footnote{\label{foot-ws4j}WS4J: \url{https://github.com/Sciss/ws4j}} project, which relies on the WordNet~\cite{wordnet} dictionary. The prototype generates a JSON format, including entities and relationships for each microservice. It also presents a list of entities that provides a holistic context map of the system after eliminating duplicated matched entities.

\noindent \textbf{Testbench:}
To demonstrate our case study, we utilized a public microservices testbench known as the train-ticket\footnote{Train-ticket V1.0.0: \url{https://github.com/FudanSELab/train-ticket/tree/v1.0.0}} testbench system. It comprises 47 microservices, with 42 of them based on the Java-based Spring Boot framework. The system adheres to enterprise conventions by employing distinct controllers, services, and repositories for layering the application. Inter-service communication between microservices in the system is facilitated through REST API calls.

\vspace{-1em}
\subsection{Results}

The prototype was executed on the testbench to construct the endpoint and data dependencies. To ensure the data extraction's completeness, the prototype outcomes were manually validated. The resulting dependencies were analyzed separately and subsequently combined to form a comprehensive dependency view of the system. The heatmap is used as the visualization approach for the dependencies. Due to space constraints, the discussion refers to the microservices IDs listed in~\tab{ms-list}. For more detailed results, please refer to the provided dataset\footnote{\label{foot-dataset}Dataset: \url{https://zenodo.org/record/8106860}}.

\begin{figure}[h!]
        \centering
        \includegraphics[width=0.85\linewidth]{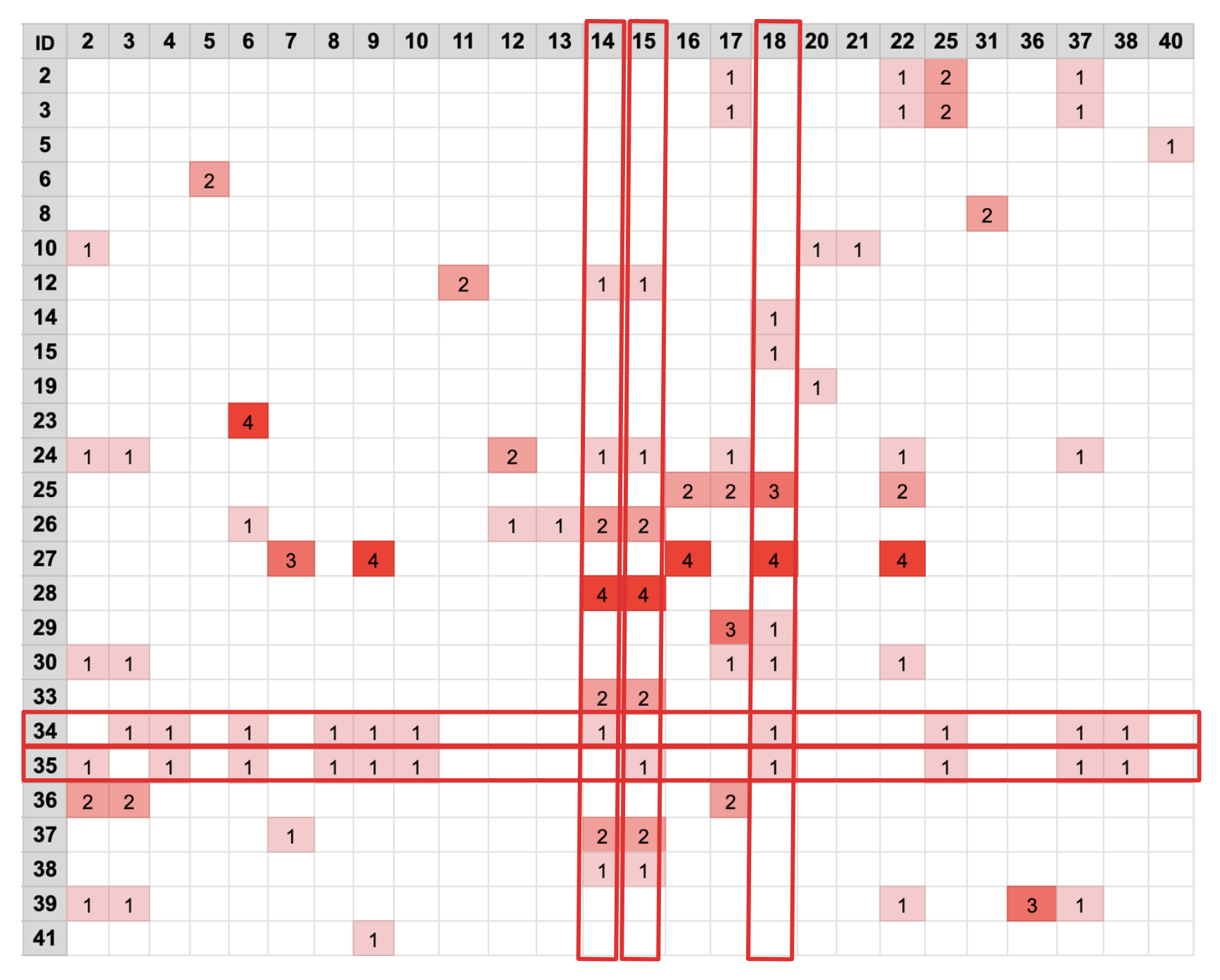}
        \vspace{-1em}
       \caption{Endpoint Dependency Matrix (EDM).}
       {The longest rows and columns are visually marked using a red rectangle.}
      \vspace{-2em}
       \label{fig:edm}
    \end{figure}

\noindent
\textbf{Endpoint Dependency:} The endpoint dependency matrix (EDM) is depicted in~\fig{edm}. The first column represents the microservices IDs containing request calls to the microservices listed in the first row. The values within each cell indicate the number of endpoint calls between each pair of microservices. Microservices containing no request calls to other microservices have been removed from the first column. This includes the following 16 microservices: $1, 4, 7, 9, 11, 13, 16-18, 20-22, 31, 32, 40,$ and $42$. Similarly, microservices that do not have any request calls made to them have been eliminated from the first row, resulting in removing the following 16 microservices: $1, 19, 23, 24, 26-30, 32-35,~39,~41,$~and~$42$.

\begin{table}[h!]
\scriptsize
\caption{List of train-ticket microservices and associated IDs}
\label{tab:ms-list}
\begin{tabular}{|r|l|r|l|r|l|}
\hline
\multicolumn{1}{|l|}{\textbf{ID}} & \textbf{Name}             & \multicolumn{1}{l|}{\textbf{ID}} & \textbf{Name}               & \multicolumn{1}{l|}{\textbf{ID}} & \textbf{Name}                \\ \hline
\textbf{1}                        & ts-common                 & \textbf{15}                      & ts-order-service            & \textbf{29}                      & ts-admin-route-service       \\ \hline
\textbf{2}                        & ts-travel-service         & \textbf{16}                      & ts-price-service            & \textbf{30}                      & ts-admin-travel-service      \\ \hline
\textbf{3}                        & ts-travel2-service        & \textbf{17}                      & ts-route-service            & \textbf{31}                      & ts-consign-price-service     \\ \hline
\textbf{4}                        & ts-assurance-service      & \textbf{18}                      & ts-station-service          & \textbf{32}                      & ts-delivery-service          \\ \hline
\textbf{5}                        & ts-auth-service           & \textbf{19}                      & ts-food-delivery-service    & \textbf{33}                      & ts-execute-service           \\ \hline
\textbf{6}                        & ts-user-service           & \textbf{20}                      & ts-station-food-service     & \textbf{34}                      & ts-preserve-other-service    \\ \hline
\textbf{7}                        & ts-config-service         & \textbf{21}                      & ts-train-food-service       & \textbf{35}                      & ts-preserve-service          \\ \hline
\textbf{8}                        & ts-consign-service        & \textbf{22}                      & ts-train-service            & \textbf{36}                      & ts-route-plan-service        \\ \hline
\textbf{9}                        & ts-contacts-service       & \textbf{23}                      & ts-admin-user-service       & \textbf{37}                      & ts-seat-service              \\ \hline
\textbf{10}                       & ts-food-service           & \textbf{24}                      & ts-rebook-service           & \textbf{38}                      & ts-security-service          \\ \hline
\textbf{11}                       & ts-payment-service        & \textbf{25}                      & ts-basic-service            & \textbf{39}                      & ts-travel-plan-service       \\ \hline
\textbf{12}                       & ts-inside-payment-service & \textbf{26}                      & ts-cancel-service           & \textbf{40}                      & ts-verification-code-service \\ \hline
\textbf{13}                       & ts-notification-service   & \textbf{27}                      & ts-admin-basic-info-service & \textbf{41}                      & ts-wait-order-service        \\ \hline
\textbf{14}                       & ts-order-other-service    & \textbf{28}                      & ts-admin-order-service      & \textbf{42}                      & ts-gateway-service           \\ \hline
\end{tabular}
\vspace{-1em}
\end{table}


The dependency matrix showcases dependencies between multiple microservices, primarily consisting of one or two endpoint calls. However, there are four dependencies with a degree of three: $25 \rightarrow 18$, $27 \rightarrow 7$, $29 \rightarrow 17$, and $39 \rightarrow 36$. Notably, these dependencies originate from different microservices. The highest degree of dependencies observed is four, which occurs in seven pairs of microservices: $23 \rightarrow 6$, $27 \rightarrow \{9,16,18,22\}$, and $28 \rightarrow \{14,15\}$. The microservice \texttt{ts-admin-basic-info-service} (ID 27) exhibits a fourth-degree dependency on four distinct microservices, while the microservice \texttt{ts-admin-order-service}~(ID 28) relies on the microservices \texttt{ts-order-other-service} (ID 14) and \texttt{ts-order-} \texttt{service}~(ID 15), each with four endpoint calls.

Examining the longest rows containing values in the matrix reveals microservices with the highest number of dependencies, indicating that they make requests to a significant number of other microservices. For instance, the \texttt{ts-rebook-} \texttt{service} (ID 24) exhibits dependencies on eight different microservices, while the longest row belongs to \texttt{ts-preserve-other-service} (ID 34) and \texttt{ts-preserve-} \texttt{service} (ID 35) with eleven dependencies. On the other hand, analyzing the longest column highlights the microservices with the most dependants, meaning they receive requests from a greater number of microservices. The matrix indicates that \texttt{ts-route-service} (ID 17) and \texttt{ts-train-service} (ID 22) have a length of seven dependent microservices. However, the longest column contains eight dependants, which are microservices with IDs 14, 15, and 18.

\begin{table}[h!]
\vspace{-2em}
\caption{Endpoints receiving more than three calls from other microservices.}
\label{tab:endpoints-dependency}
\begin{tabular}{|c|l|l|c|c|}
\hline
\textbf{ID} & \multicolumn{1}{c|}{\textbf{Endpoint Path}}          & \multicolumn{1}{c|}{\textbf{Method}} & \textbf{\#Calls} & \textbf{$\#\mu s$} \\ \hline
\textbf{17} & ts-route-service/api/v1/routeservice/routes          & GET                                  & 8                & 7             \\ \hline
\textbf{18} & ts-station-service/api/v1/stationservice/stations/id & GET                                  & 4                & 3             \\ \hline
\textbf{22} & ts-train-service/api/v1/trainservice/trains/byName   & GET                                  & 6                & 6             \\ \hline
\textbf{25} & ts-basic-service/api/v1/basicservice/basic/travel    & POST                                 & 6                & 4             \\ \hline
\end{tabular}
\vspace{-2em}
\end{table}
    
Further analysis delves into whether the dependants of a microservice make requests to the same endpoint or if they are spread across multiple endpoints within the microservice. The table presented in~\tab{endpoints-dependency} highlights the endpoints that receive multiple requests from other microservices, specifically focusing on endpoints with more than three requests. It is important to note that not every call originates from a distinct microservice as shown in column ($\#\mu s$). Notably, the GET endpoint with the path \texttt{ts-route-service/api/v1/routeservice/route} receives eight calls from seven different microservices. This observation could indicate a potential functional bottleneck in the system, where multiple microservices rely on this endpoint to fulfill their respective use cases.

\begin{figure}[h!]
\vspace{-2em}
        \centering
        \includegraphics[width=0.85\linewidth]{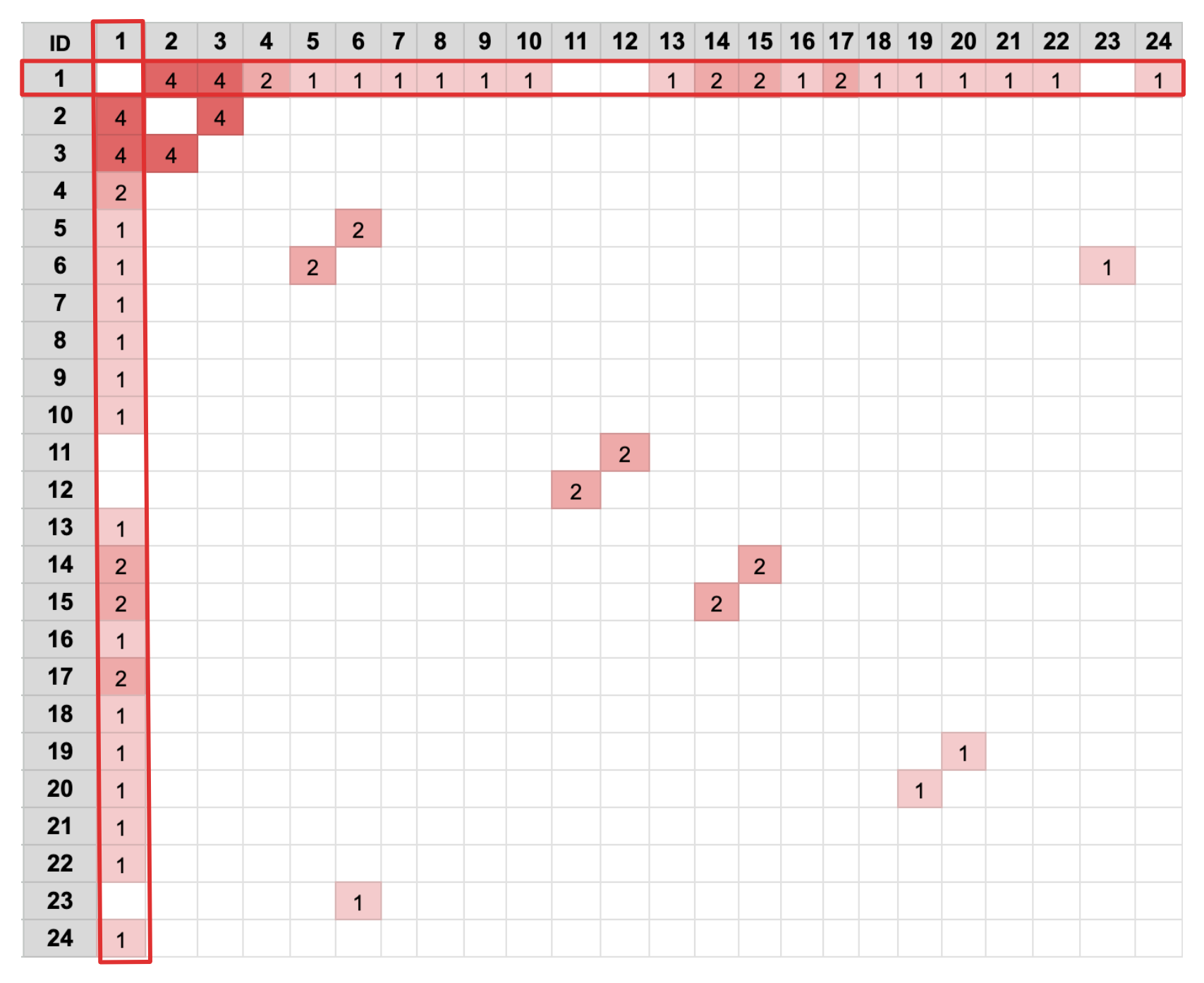}
        \vspace{-1em}
       \caption{Data Dependency Matrix (DDM).}
       {The longest rows and columns are visually marked using a red rectangle.}
      \vspace{-2em}
       \label{fig:ddm}
    \end{figure}

\noindent
\textbf{Data Dependency:} The data dependency matrix (DDM) in~\fig{ddm} represents the number of common data entities between microservice pairs. The rows and columns correspond to microservice IDs, while the cell values indicate the count of matched data entities. Unlike the endpoint dependency matrix (EDM), this matrix is symmetric and undirected, meaning the values remain the same regardless of whether one starts from the rows or columns. A total of 18 microservices (IDs 25-42) have been excluded from the rows and columns of the DDM because they do not share any common data entities with other microservices.

The matrix reveals that multiple microservices share one or two common data entities with other microservices. However, the maximum number of common entities between a pair of microservices is four, observed between \texttt{ts-common} (ID 1) and both \texttt{ts-travel-service} (ID 2) and \texttt{ts-travel2-service} (ID 3), and also between \texttt{ts-travel-service} (ID 2) and \texttt{ts-travel2-service} (ID 3).

Moreover, the longest row in terms of values belongs to \texttt{ts-common} (ID 1), indicating that this microservice shares the most common entities with other twenty microservices. However, the next longest row corresponds to \texttt{ts-user-service} (ID 6) with a length of only three, highlighting a significant disparity in data dependencies among the microservices, with a concentration of dependencies in a single microservice (\texttt{ts-common}). Upon further examination of the most common data entities across all microservices, we identified eight commonly shared entities: \texttt{AdminTrip, Order, OrderAlterInfo, StationFoodStore, Travel, Trip, TripAllDetail}, and \texttt{User}. All these entities also exist in \texttt{ts-common}, but are shared only across three distinct microservices.



\begin{figure}[b!]
\vspace{-2em}
        \centering
        \includegraphics[width=0.85\linewidth]{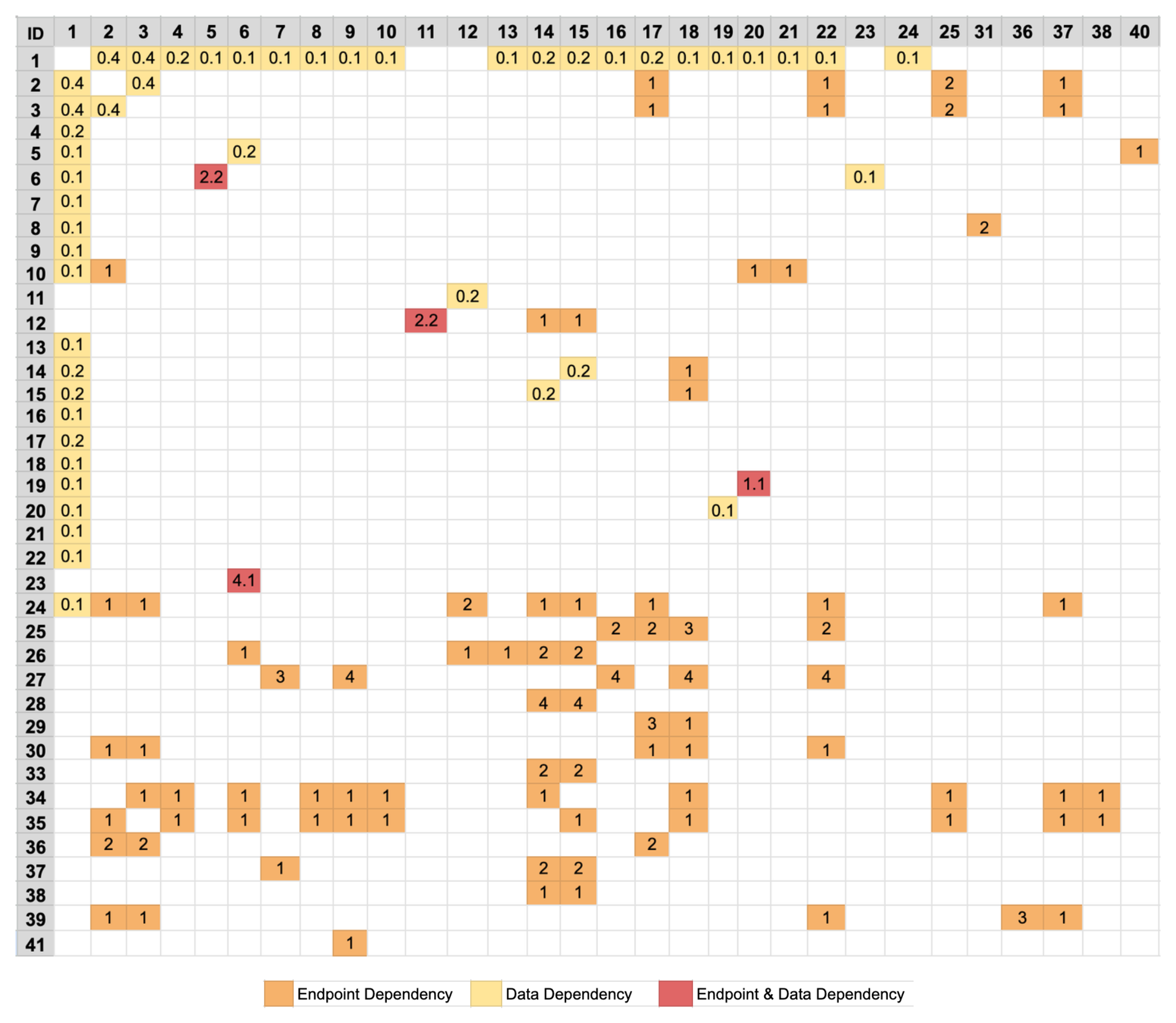}
        \vspace{-1em}
       \caption{Service Dependency Matrix (SDM).}
      \vspace{-2em}
       \label{fig:sdm}
    \end{figure}

\noindent
\textbf{Comprehensive Service Dependency:} By combining the EDM and the DDM, we generate a comprehensive perspective of the system's dependencies known as the Service Dependency Matrix (SDM), as shown in~\fig{sdm}. The SDM represents microservice IDs as both columns and rows. The cell values in the SDM are decimal numbers, where the integer part corresponds to the endpoint dependency degree from the EDM, and the fractional part corresponds to the data dependency degree from the DDM. 

To visually distinguish between different types of dependencies, the matrix utilizes different colors for endpoints-only dependencies, data-only dependencies, and dependencies involving both endpoints and data. The inclusion of data dependency in the fractional part of the SDM does not diminish its value compared to the endpoint dependencies. The construction of the decimal value is primarily related to the data formatting rather than the absolute significance of the cell value. For instance, consider the cell at position (row: 23, column: 6) in the SDM, which has a value of 4.1. This value indicates that \texttt{ts-admin-user-service} microservice (ID 23) has made four calls to \texttt{ts-user-service} microservice (ID 6), and there is one common entity (\texttt{UserDto}) shared between them.

Analyzing the SDM, it becomes apparent that the responsibility of holding common data entities among microservices is predominantly concentrated in the \texttt{ts-common} microservice. This concentration results in distinct separations between the dependencies of endpoints and data entities. However, some overlaps can still be observed between the following four microservice pairs: $6 \rightarrow 5$, $12 \rightarrow 11$, $19 \rightarrow 20$, and $23 \rightarrow 6$. These pairs demonstrate a strong dependency within the system, as they depend on each other for both direct endpoint calls and the presence of common data entities. These dependencies highlight their interconnected nature and the importance of their mutual interaction.

\section{Discussion} \label{sec:discussion}






In the proposed method, we aim to provide a comprehensive understanding of system dependencies by considering both the endpoints and data perspectives. The introduced dependency matrices present system-centric perspectives that have the potential to provide a scalable visualization approach, helping practitioners in comprehending the system architecture and its dependencies. Blending endpoint dependencies (EDM) and data dependencies (DDM) within a unified matrix (SDM) has the potential to unveil more profound architectural concerns within microservices applications, surpassing what can be discerned from the separate EDM and DDM matrices. Moreover, by comparing the metrics across different versions, we can track the evolution of system dependencies over time.

While our method and prototype are valuable, it is important to acknowledge their limitations, particularly regarding the consideration of other perspectives of dependencies. The asynchronous communication model between microservices (e.g., publish-subscribe pattern), is not currently covered by our approach and they are not used in the train-ticket testbench as well. Incorporating such perspectives would provide additional insights into the interconnections between system components beyond the direct endpoint calls. Furthermore, this study focuses on analyzing the system's source code to gain a holistic understanding of all possible execution paths. However, considering the runtime interactions captured in logs and traces could provide valuable insights into the actual number of calls made to a particular microservice. This additional perspective could offer an additional depiction of the dependencies between microservices and enhance our understanding of the system's behavior.

\noindent
\textbf{Threats to Validity:}
The method does not address all potential microservice dependencies, its purpose is to illustrate how dependency matrices can assist in system analysis. Our prototype tool is tailored for the Java platform, potentially restricting its relevance to other programming languages. However, it's important to emphasize that the focus was on introducing the methodology rather than creating an exhaustive tool. 
%
%
In certain cases, the prototype tool might encounter challenges in accurately matching method signatures, particularly in situations where there are ambiguous method names. Additionally, the entity matching process is currently restricted to basic similarities such as names and field matches, indicating that there are inherent limitations in approximation.

The case study analysis may be influenced by specific constructs present in the selected testbench, potentially limiting the prototype's generalizability across different systems. However, manual validation of the prototype's outcomes was performed to ensure the completeness of information extraction from the source code. Furthermore, the chosen testbench is employed in various research and is regarded as a well-established and representative microservice system.

\section{Related Work} \label{sec:rw}




Numerous studies underscore the significance of managing dependencies in microservice architectures. According to Lewis and Fowler~\cite{microservice2014lewis}, loosely coupled microservices offer advantages in independent modifications but pose challenges as systems evolve. To analyze such dependencies, scholars have introduced various techniques. Apolinário et al.\cite{apolinario2021method} focus on endpoint calls, Sangal et al.\cite{sangal2005using} employ static analysis for dependency models, and Eski and Buzluca~\cite{eski2018automatic} use evolutionary code coupling. Our approach uses static analysis to extract and integrate both endpoint and data dependencies for a comprehensive system view.


In the realm of heterogeneous dependencies in distributed systems, Fang et al.~\cite{10092708} devised specialized tools for compile-time dependency extraction through static analysis. They targeted entity dependencies within components and hard-coded API dependencies, using text comparison. In contrast, our method goes beyond text-based analysis, incorporating semantic similarities and fine-grained dependency capture through signature matching.


Effective visualization is crucial for comprehending system dependencies. Multiple studies \cite{rahman2019curated,oberhauser2019vr,cerny2022microvision} propose graph-based visualizations depicting microservice dependencies, focusing on communication patterns via endpoint calls. In contrast, our approach employs dependency matrices to visualize and analyze the system, offering a distinct view of microservices' dependencies.


\section{Conclusion} \label{sec:conclusion}

System dependency analysis in microservices provides valuable insights for practitioners to comprehend the system. This paper integrates endpoint and data dependencies, offering a comprehensive understanding of system dependencies and facilitating informed decision-making in developing and evolving microservice-based systems. The analysis is addressed through static code analysis providing perspectives that enable reasoning about system maintainability and monitoring system dependency evolution across different versions.
Our approach encompasses a detailed analysis of individual microservices, combining the results to a holistic dependency perspectives that can be visualized and interpreted. The focus was on generating the EDM and DDM from the source code and further combining them to create the SDM for a more comprehensive perspective. The proposed methodology was implemented in a prototype and validated through a case study, highlighting its efficacy in understanding system~dependencies.
 
Future work will include asynchronous call dependencies, recognizing their importance. We also aim to expand the prototype for analyzing system polyglots.

\vspace{-0.5em}
\section*{Acknowledgements}
\vspace{-6px}
This material is based upon work supported by the National Science Foundation under Grant No. 2245287.

\vspace{-.5em}



\bibliographystyle{splncs04}
\bibliography{references}
%






\end{document}